\documentclass[journal=aamick,manuscript=article, layout=draft]{achemso}

\usepackage{balance}
\usepackage{mathptmx}
\usepackage{sectsty}
\usepackage{graphicx}
\usepackage{lastpage}
\usepackage{caption}

\usepackage{epstopdf}

\usepackage{fancyhdr}
\usepackage{fnpos}
\usepackage[english]{babel}
\addto{\captionsenglish}{%
  
}
\usepackage{array}
\usepackage{droidsans}
\usepackage{charter}
\usepackage[T1]{fontenc}
\usepackage[usenames,dvipsnames]{xcolor}
\usepackage{setspace}
\usepackage[compact]{titlesec}

\usepackage{float}
\usepackage{ifthen}
\usepackage{textcomp}
\usepackage{overpic}
\usepackage{multirow}
\usepackage{xr}
\makeatletter
\newcommand*{\addFileDependency}[1]{
  \typeout{(#1)}
  \@addtofilelist{#1}
  \IfFileExists{#1}{}{\typeout{No file #1.}}
}
\makeatother

\newcommand*{\myexternaldocument}[1]{
    \externaldocument{#1}
    \addFileDependency{#1.tex}
    \addFileDependency{#1.aux}
}
\myexternaldocument{Supplement}

\usepackage{bbm}
\usepackage{tabularx}
\usepackage{amsmath}
\usepackage{amssymb}
\usepackage{gensymb}
\usepackage{pifont}

\usepackage{color}

\newcommand{\mahg}{mAh g$^{-1}$}

\newcommand{\LiFeSeO}{(Li$_2$Fe)SeO}
\newcommand{\lise}{(Li$_2$Fe)SeO}
\newcommand{\lis}{(Li$_2$Fe)SO}
\newcommand{\LiFeSO}{(Li$_2$Fe)SO}
\newcommand{\LiFeSexSyO}{(Li$_2$Fe)S$_{1-x}$Se$_x$O}

\newcommand{\etal}{\textit{et~al.}}

\newcommand{\feS}{Fe$_{\mathrm{1-x}}$S$_{\mathrm{x}}$}

\usepackage{array}
\usepackage[version=3]{mhchem} %

\author{L.~Singer}
\affiliation{Kirchhoff Institute for Physics, Heidelberg University, 69120, Heidelberg, Germany}
\email{lennart.singer@kip.uni-heidelberg.de}
\author{Bowen Dong}
\affiliation{Kirchhoff Institute for Physics, Heidelberg University, 69120, Heidelberg, Germany}
\author{M.A.A. Mohamed}
\affiliation{Leibniz Institute for Solid State and Materials research (IFW) Dresden, 01069 Dresden, Germany}
\alsoaffiliation{Department of Physics, Faculty of Science, Sohag University, 82524 Sohag, Egypt}

\author{F. L. Carstens}
\affiliation{Kirchhoff Institute for Physics, Heidelberg University, 69120, Heidelberg, Germany}

\author{Silke Hampel}
\affiliation{Leibniz Institute for Solid State and Materials research (IFW) Dresden, 01069 Dresden, Germany}

\author{N. Gr\"a\ss ler}
\affiliation{Leibniz Institute for Solid State and Materials research (IFW) Dresden, 01069 Dresden, Germany}

\author{R. Klingeler}
\affiliation{Kirchhoff Institute for Physics, Heidelberg University, 69120, Heidelberg, Germany}
\email{klingeler@kip.uni-heidelberg.de}

\title{{Separating cationic and anionic redox activity in antiperovskite \LiFeSO}}

\begin{document}




\begin{abstract}
Lithium-rich antiperovskite promise to be a compelling high-capacity cathode material due to existence of both cationic and anionic redox activity. Little is however known about the effect of separating the electrochemical cationic from the anionic process and the associated implications on the electrochemical performance.
In this context, we report the electrochemical properties of the illustrative example of three different \LiFeSO\ materials with a focus on separating cationic from anionic effects. With the high voltage anionic process, an astonishing electrochemical capacity of around 400~\mahg\ can initially be reached.  Our results however identify the anionic process as the cause of poor cycling stability and demonstrate that fading reported in previous literature is avoided by restricting to only the cationic processes.  Following this path, our \LiFeSO-BM500 shows strongly improved performance indicated by constant electrochemical cycling over 100 cycles at a capacity of around 175~\mahg\ at 1~C. Our approach also allows us to investigate the electrochemical performance of the bare antiperovskite phase excluding extrinsic activity from initial or cycling-induced impurity phases. Our results underscore that synthesis conditions are a critical determinant of electrochemical performance in lithium-rich antiperovskites, especially with regard to the amount of electrochemical secondary phases, while the particle size has not been found a crucial parameter. Overall, separating and understanding the effects of cationic from anionic redox activity in lithium-rich antiperovskites provides the route to further improve their performance in electrochemical energy storage.
\end{abstract}

\section{Introduction}
The recent discovery of the lithium-rich antiperovskite compounds characterized by the general formula (Li$_2$TM)ChO (with TM = Fe, Mn, Co; Ch = S, Se) has significantly broadened the landscape of potential cathode materials for lithium-ion batteries (LIBs)~\cite{Lai.2017}. This novel class of materials showcases highly favorable attributes in the context of lithium-ion battery application, including cost-effectiveness, utilization of environmentally benign raw materials, efficient lithium diffusion, and the ability of multi-electron storage per chemical unit~\cite{Assat.2018-anionic,Mikhailova.2018,Lu.2018,Song.2017-multi,Cui.2023-multielectron,Deng.2022, Dai.2024}.
From the so far investigated materials Li$_2$FeSO \cite{Lai.2017,Lu.2018,Gorbunov.2020,Gorbunov.2021-frontiers,Miura.2022,Deng.2023-Li2FeMnSO,Mohamed.2023-synthese,singer2023elucidating}, (Li$_2$Co)SO\cite{Lai.2018-flexible}, (Li$_2$Mn)SO \cite{Lai.2018-flexible}, (Li$_2$Fe$_{1-x}$Mn$_{x}$)SO\cite{Gorbunov.2020,Gorbunov.2021-frontiers,Deng.2023-Li2FeMnSO}, (Li$_2$Fe$_{0.9}$Co$_{0.1}$)SO\cite{Gorbunov.2021-frontiers}, (Li$_2$Co)SeO\cite{Lai.2018-flexible}, (Li$_2$Mn)SeO\cite{Lai.2018-flexible}, \LiFeSexSyO\ \cite{Mohamed.2021} and \LiFeSeO\ \cite{Lai.2017,Li2FeSeO-2022,singer2023elucidating}, the (Li$_2$Fe)SO compound captivates with the highest theoretical capacity. Despite the promises of lithium-rich antiperovskites and especially \LiFeSO, their great potential could not be fully exploited due to poor cycling stability (only 66~\% capacity retention in \LiFeSO\ at 0.1~C after 50 cycles) \cite{Mikhailova.2018}.

In our recent works on \LiFeSeO, we have demonstrated the stabilizing effect of selenium as an anion in the antiperovskite structure for the electrochemical performance \cite{Li2FeSeO-2022}. We have also shown that poor cycling stability in this system originates from the high-voltage electrochemical processes causing decomposition of the antiperovskite structure \cite{singer2023elucidating}. When cutting off the decomposition reaction, the cycle stability is significantly improved (over 80~\% capacity retention after 100 cycles for \LiFeSeO\ at 1~C in the voltage range from 1 to 2.5~V)~\cite{singer2023elucidating}.
A big challenge to further improve cycling stability in \LiFeSeO\ is however the clear separation of the second step of the cationic two-stage iron oxidation reaction from the highly overlapping anionic decomposition reaction.
Studies conducted on the \LiFeSexSyO\ series have unveiled a noteworthy phenomenon to solve this issue: An increase in sulfur content (from $x=0.1$ to 0.9) correlates with a discernible upward shift in the voltage range corresponding to the anionic oxidation reaction while the cationic iron reaction remains essentially invariant~\cite{Mohamed.2021}. This empirical observation offers a route to disentangle and hence separately investigate the cationic and suggested anionic reactions by studying the S-end member of the doping series which manifests the most distinct voltage difference between the cationic and anionic reactions.

Building on these results, we present the electrochemical properties of three differently mechanochemically synthesized \LiFeSO\ samples and explore the effect of separating the electrochemical cationic from the anionic process. Following this path demonstrates that the issue of cycle stability clearly can be improved. Optimised \LiFeSO\ exhibits a high capacity of around 175~\mahg\ which only marginally decreases by less than 4~\% over 100 cycles at 1C if the electrochemical processes are restrained to only the cationic processes. Including the anionic processes yields after a few cycles twice this capacity, but severe fading by more than 75~\% (typical for Li-rich antiperovskites~\cite{Mikhailova.2018,Gorbunov.2020,singer2023elucidating}), displays that this process is detrimental for cycling stability. Our results further unveil that particle size is not a key parameter for electrochemical performance in Li-rich antiperovskites, but impurities present in the samples govern the observed differences.

\section{Experimental section}

\subsection*{Synthesis and structural characterisation}

The synthesis employed, as well as a detailed physical characterization of the \LiFeSO\ samples, can be found in Ref.~\cite{Mohamed.2023-synthese}. The \LiFeSO\ samples used here have been produced by a ball milling (BM) method. Either pristine material was used or post-synthesis heat-treatment at 300~\degree C or 500~\degree C was applied.\footnote{The materials under study have been labelled 'BM', '300~\degree C', and '500~\degree C' in Ref.~\cite{Mohamed.2023-synthese}.} To highlight the synthesis conditions, the samples will henceforth be referred to as BM, BM300, and BM500.
X-ray diffraction studies have been performed  (STOE STADI P) in Debye-Scherrer mode with Mo K$\alpha_1$ radiation source ($\lambda$ = 0.70926 \AA) and a Mythen 1~K detector (Dectris). To prevent any air exposure the samples were filled into glass capillaries inside the argon glove box and then melt-sealed.


\subsection*{Magnetic measurements}
Magnetic measurements were performed on \lis\ powder samples using a MPMS3 magnetometer (Quantum Design). The static magnetic susceptibility $\chi =M/B$ has been obtained upon varying the temperature between 120 and 350~K at 1~T by using field-cooled (FC) and zero-field-cooled (ZFC) protocols where the sample was cooled either in the external measurement field or the field was applied after cooling to the lowest temperature. Isothermal magnetisation $M(B)$ has been obtained at $T=1.8$~K in the field range -7~T~$\leq B\leq$~+7~T.

\subsubsection*{Electrochemical measurements}
All electrode and battery assembly has been performed in an Argon-filled glovebox with controlled humidity and oxygen concentration due to the moisture sensitivity of \LiFeSO. The electrode slurry was prepared by mixing the active material, carbon black, and polyvinylidene fluoride with the ratio of 70:15:15 (wt.\%) in anhydrous isopropanol. Afterwards, the resulting mixture was spread onto a (\O=10~mm) aluminum mesh current collector to manufacture electrodes with an active material mass loading between 3.2 and 6.3~mg per cm$^{-2}$ (see \cite{Ottmann.2015}). The obtained electrodes were thereafter dried overnight at room temperature and then pressed at around 7.5~MPa. As separator Glass fiber (Whatman GF/D) and as the counter electrode, pure lithium metal foil were used \cite{Singer.2023}. The electrolyte is composed of 1M LiPF$_6$ in a mixture of ethylene carbonate and dimethyl carbonate (1:1 by volume, Aldrich). Electrochemical measurements were carried out using a VMP3 potentiostat and a BCS-805 (BioLogic) at 25$~\degree \mathrm{C}$.
Cyclic voltammetry (CV) and galvanostatic cycle measurements with potential limitation (GCPL) were performed in a 2023 coin cell housing~\cite{Zakharova.2021}.
Note, the theoretical capacity of \lis\ for the extraction/insertion of 1~Li$^+$ is 227.57~\mahg\ and the rate 1~C is defined as the charge/discharge current required to extract/insert 1~Li$^+$ from/into the antiperovskite material within one hour.

\section{Results and discussion}

\begin{figure}[htb]
    \centering
    \includegraphics[width=\columnwidth]{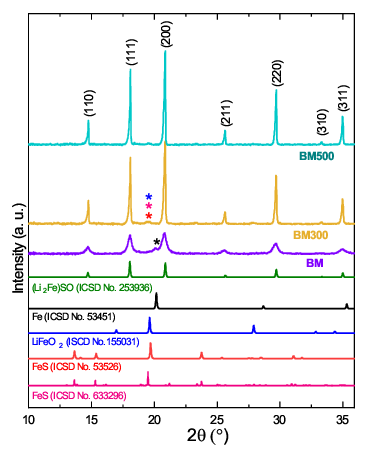}
    \caption{XRD patterns of BM, BM300 and BM500 as well as reference patterns for \lis\ (ICSD No. 253936 \cite{Lai.2017}), Fe (ICSD No. 53451 \cite{Fe-Ref.1967}),  LiFeO$_2$ (ICSD No. 155031 \cite{VucinicVasic.2006-LiFeO2-Ref}) and FeS (ICSD No. 53526 \cite{Coey.1979-FeS-53526} and 6332969 \cite{Morimoto.1970-FeS-63}).}
    \label{fig:XRD}
\end{figure}

\begin{figure}[htb] \centering
  \includegraphics[width=\columnwidth]{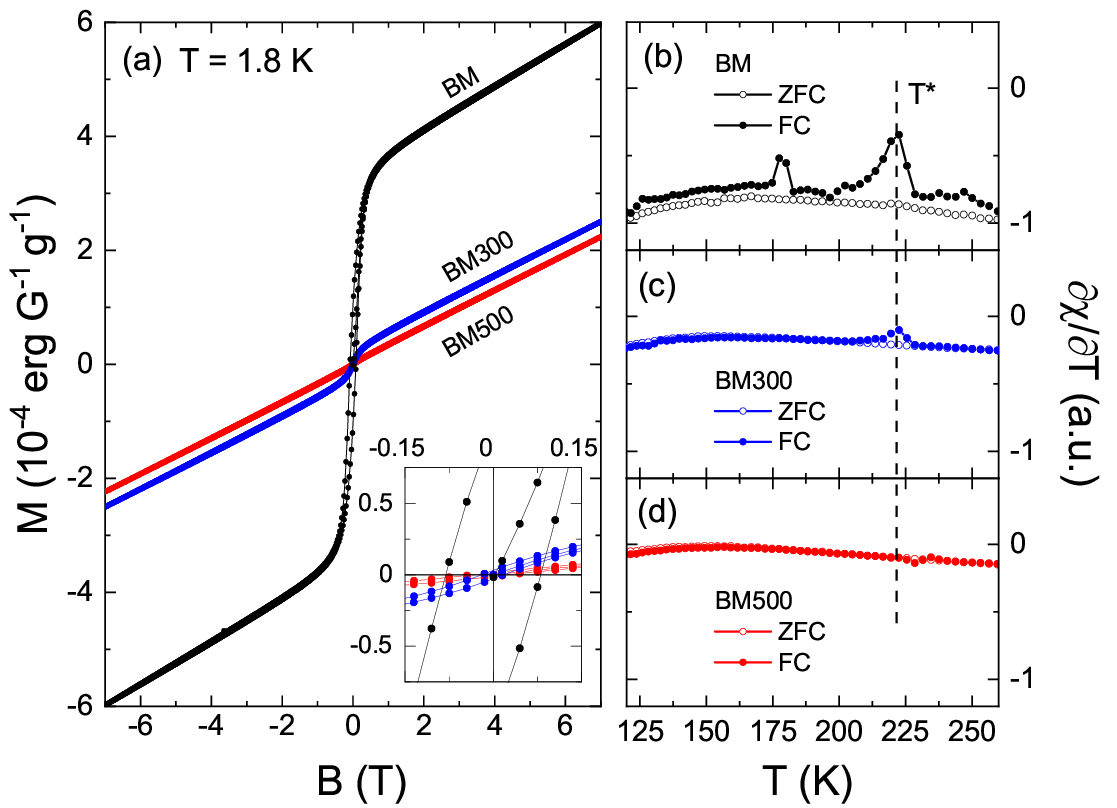}
    \caption{(a) Isothermal magnetisation of BM, BM300 and BM500 vs. magnetic field, at 1.8~K. Temperature derivative of the static magnetic susceptibility ($\chi=M/B$) of the BM (b), BM300 (c) and BM500 (d) sample vs. temperature at B = 1~T in the temperature range between 120 and 260~K. Both FC (orange) and ZFC (grey) protocols have been applied. The anomaly labeled by $T^*$ signals a tiny Fe$_{0.95}$S impurity phase, which strongly decreases for BM300 and vanishes for BM500.}
    \label{fig::Mag}
\end{figure}

Figure~\ref{fig:XRD} displays powder XRD patterns of the mechanochemical synthesized \lis-BM, \lis-BM300, and \lis-BM500 together with the reference patterns of \lis\ (ICSD No. 253936 \cite{Lai.2017}) and some possible impurity phases: Fe (ICSD No. 53451 \cite{Fe-Ref.1967}), LiFeO$_2$ (ICSD No. 155031 \cite{VucinicVasic.2006-LiFeO2-Ref}) and FeS (ICSD No. 53526 \cite{Coey.1979-FeS-53526} and 6332969 \cite{Morimoto.1970-FeS-63}). The cubic antiperovskite structure with  {\it Pm$\overline{3}$m} space group is clearly the main phase for all three materials. Small additional peaks appearing mostly between 19 and 20$\degree$ in the patterns of the \lis\ samples are associated with the presence of crystalline impurities. We refer to Figure.~\ref{si-fig::XRD} in the Supplementary Information for a closer look at the patterns. The data imply that the BM material contains an iron impurity phase. After subsequent heat treatment (BM300, BM500) this phase is however no longer present. All samples show a small elevation around 19.5~$\degree$ which may indicate either FeS and/or LiFeO$_2$. Given the very similar reflexes at around 19.5$\degree$ of FeS and LiFeO$_2$, it is not possible to make a conclusive determination based on the XRD data alone.

The fact that many iron-based materials show long-range magnetic order renders measurements of the magnetisation a particularly suitable tool to determine and quantify potential foreign phases with exceptionally high sensitivity~\cite{Chernova.2011-magnetism}. Figure~\ref{fig::Mag}a depicts the isothermal field-dependent magnetization $M$($B$) at 1.8~K for all three materials under study. While in all three materials the magnetic susceptibility $\partial M/\partial B(|B|\gtrsim 1~{\rm T})$ is very similar, there is a pronounced difference around zero field. Specifically, there is a large jump in magnetisation for BM which is strongly reduced in BM300 and nearly vanishes for BM500. The data imply a ferromagnetc moment of $M_{\rm s}$ = 3.3(1)~erg G$^{-1}$g$^{-1}$, 0.23(1)~erg G$^{-1}$g$^{-1}$, and 0.05(5)~erg G$^{-1}$g$^{-1}$ for BM, BM300, and BM500, respectively. Hence, while BM shows a well-visible ferromagnetic impurity phase, it decreases by one respectively two orders of magnitude upon heat treatment at 300 and 500~$^\circ$C. By attributing the ferromagnetic signal to an iron impurity phase, comparison of $M_{\rm s}$ with the saturation magnetization of iron $M^{\mathrm{Fe}}_{\mathrm{s}}$ = 220.25~erg G$^{-1}$ g$^{-1}$ (see Ref.~\cite{Bozorth.1978-Fe}) yields a contained iron impurity of maximally 1.4 wt\%. This value agrees well with 1.2 vol\% for BM inferred from XRD data~\cite{Mohamed.2023-synthese}. We note however, that $M_{\rm s}$ can contain contributions of other ferro- or ferrimagnetic impurity phases like \feS\ or LiFeO$_2$ which can not be ruled out either.

The temperature dependence of the temperature derivative $\partial \chi/\partial T$ of the pristine material BM (Fig.~\ref{fig::Mag}b) confirms the presence of an additional impurity phase. Specifically, the FC data displays an anomaly at $T^*\simeq 222$~K, indicative for an antiferromagnetic transition which we attribute to the alpha transition in Fe$_{0.95}$S~\cite{Adachi.1988-FeS}. The associated anomaly at $T^*$ strongly decreases for BM300 and vanishes for BM500 (see Figs. \ref{fig::Mag}b to d ).

\begin{figure}[htb]
    \centering
    \includegraphics[width=0.5\columnwidth]{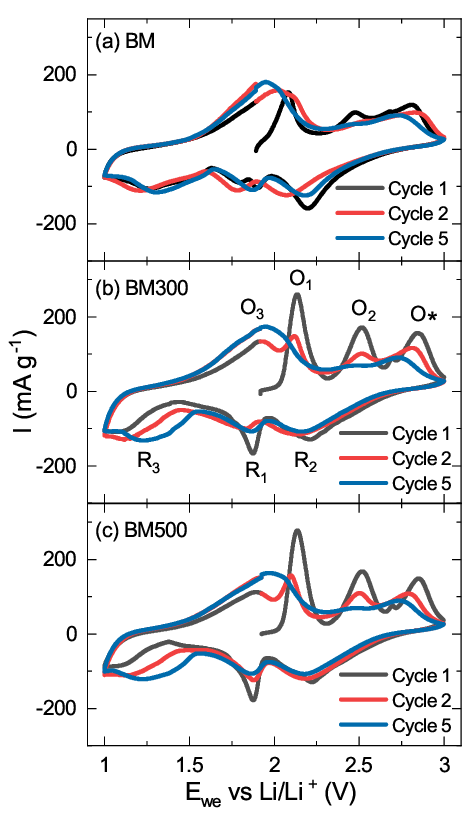}
    \caption{Cyclic voltammograms of cycle 1, 2 and 5 of (a) BM (b) BM300 and (c) BM500 measured at a scan rate of 0.1~mV~s$^{-1}$. The labels R/O mark distinct reduction and oxidation features.}
    \label{fig:full-CV}
\end{figure}

\begin{figure*}[htb] \centering
  \includegraphics[width=0.9\textwidth]{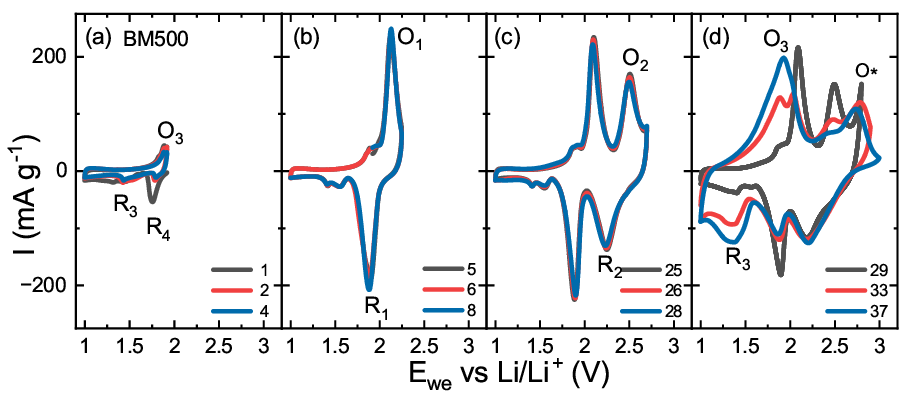}
    \caption{Cyclic voltammograms of BM500 in the potential range from 1~V  to (a) OCV (b) 2.3 (c) 2.7~V (d) to 2.8~V (cycle 29) to 2.9 (cycle 33) and  3~V (cycle 37)  vs. Li/Li$^+$ measured at a scan rate of 0.1~mV~s$^{-1}$. The labels R/O correspond to the ones in Fig.~\ref{fig:full-CV}.}
    \label{fig::Cv-Bm500}
\end{figure*}

Cyclic voltammetry is used to investigate the electrochemical processes in BM, BM300, and BM500 in detail.
Notably, the cyclic voltammograms (CVs) of all three \lis\ materials in Fig.~\ref{fig:full-CV} demonstrate a comparable multistage lithium extraction and insertion mechanism. Independent on the synthesis conditions, the two initial oxidation peaks O$_1$ and O$_2$ appear at about 2.1~V  and 2.5~V and signal the two-stage cationic Fe$^{2+}$ to Fe$^{3+}$ oxidation process \cite{Mikhailova.2018}. The high-voltage oxidation peak O$*$ at 2.8~V is believed to be originating from an anionic reaction~\cite{Mikhailova.2018}.
During the first intercalation (discharge), all three samples display distinct reduction peaks (R$_1$, R$_2$ and R$_3$). This observation is consistent with previous research on solid-state reaction (SSR) synthesized \lis~\cite{Mikhailova.2018,Gorbunov.2021-frontiers}.
The insertion and removal of lithium via multiple stages moreover is observed in all lithium-rich antiperovskites investigated so far and can therefore be assumed to be a general feature of this material class. The same is true for the decrease of the initial differences in the electrochemical properties of the different materials after only a few cycles.
Based on our previous work on \lise~\cite{singer2023elucidating}, where we showed that the high-voltage anionic processes are crucial for the electrochemical performance and properties, CV  measurements were performed for the three \lis\ materials under study where the maximum voltage is step-wise increased (see Fig.~\ref{fig::Cv-Bm500} for BM500 and Fig.~\ref{si-fig::cv-all} in the Supplemental Material for BM and BM300). Here, a special focus was placed on the separation of cationic and anionic redox activity.

CV measurements with initial lithiation (discharge) to 1~V and the restriction of the maximum voltage to OCV (Fig.~\ref{fig::Cv-Bm500}a and Fig.~\ref{si-fig::cv-all}) show the electrochemical behavior at initial intercalation. Restricting the voltage only to the regime of lithium insertion provides the electrochemical activity of possible lithium vacancies or active anodic impurity phases. The data in Figs.~\ref{fig::Cv-Bm500}a and \ref{si-fig::cv-all}a-c reveal electrochemical activity signalled by R$_3$ at approx. 1.4~V and O$_3$ at 2~V, which originates from the following conversion reaction of  \feS\ \cite{Fei.2013-FeS,Liu.2018-FeS,Cao.2019-FeS,Xu.2018-iron-sulfides}:
\begin{equation}
    \mathrm{FeS} + 2 \mathrm{Li}^++2\mathrm{e}^- \leftrightarrow \mathrm{Li_2S}+\mathrm{Fe}
    \end{equation}

The comparison between the three different materials displays that the electrochemical fingerprints of \feS\ strongly decrease in the post-synthesis heat-treated materials as evidenced by the decrease of the R$_3$/O$_3$ redox peaks. This is in accordance with our magnetization data, which also show the strong decrease of the contained Fe$_{0.95}$S phase in the heat-treated materials BM300/BM500. The origin of peak R$_4$ is unclear and needs further investigation. From the fact that R$_4$ disappears after the first cycle in all samples, we conclude that either the corresponding oxidation process appears outside the designated potential window or that R$_4$ corresponds to an irreversible process. The prominence of R$_4$ increases from the BM to the BM300 sample. For the BM500 sample, R$_4$ is in the first cycle at higher voltages and less pronounced as compared to BM300.

\begin{figure*}
\begin{center}
 \includegraphics[width=\textwidth]{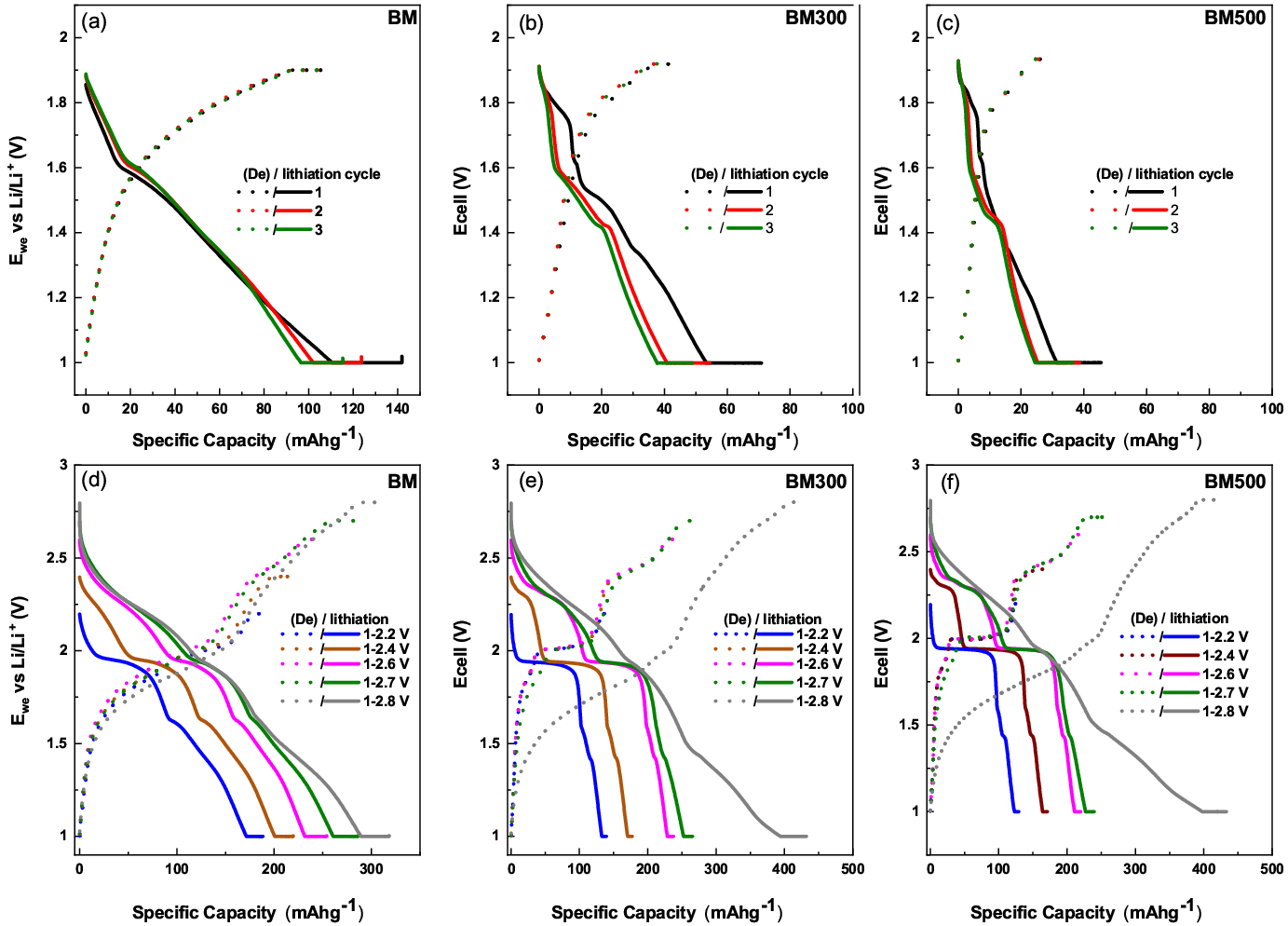}
    \caption{Potential profiles of BM (a), BM300 (b), BM500 (c) by direct discharge. Selected potential profiles of BM (d), BM300 (e), BM500 (f)  measured by a constant current constant voltage protocol (I=C/10 and a holding time of 5~h) and a step-like increase in 0.1 V increments of the maximum voltage. The colors used symbolize distinct voltage regions. }
   \label{si-fig::stepgcpl}
   \end{center}
\end{figure*}

Complementary GCPL measurements provide further insights into the extent of charge conversion (see Figs.~\ref{si-fig::stepgcpl}a-c). The data reveal a substantial conversion of charges of the contained impurity phases of approximately 100~\mahg\ for BM, around 45~\mahg\ for BM300 and 20~\mahg\ for BM500.
An estimate based on the fraction of \feS\ in the BM sample by comparing the converted capacity at R$_3$ of approximately 70~\mahg\ with the theoretical capacity of FeS (609~\mahg) \cite{Xu.2018-iron-sulfides} and FeS$_2$ (894~\mahg) \cite{Xu.2018-iron-sulfides} leads to a possible content between 11 to 7\% of \feS\ in the pristine BM material. For BM300 the capacity associated with R$_3$ is around 30~\mahg\, which leads to an estimated content between 5 to 3\% and for BM500 with only around 20~\mahg\ to a content between 3 to 2\%. Note that this this calculation tends to underestimate, as a conversion of the full theoretical capacity is rather unlikely.

Increasing the upper potential limit (see Fig.\ref{fig::Cv-Bm500}b and c) allows us to comprehensively study the step-wise redox dynamics involving the Fe$^{2+}$ $\leftrightarrow$ Fe$^{3+}$ pair. Two redox pairs successively appear when increasing the voltage regime up to 2.7~V which we attribute to Fe$^{2+}$/Fe$^{3+}$ redox processes of the antiperovskite phase~\cite{Mikhailova.2018}. In consecutive cycles, the CV curves nearly perfectly overlap, which implies highly reversible and stable processes. When extending the CVs to the full potential range of 1 to 3~V and thereby including the high-voltage anionic process O$_*$  as shown in Fig.~\ref{fig::Cv-Bm500}d, all samples show a significant change of the CV curves with a pronounced non-reversible element. Specifically, the pair R$_{3}$/O$_3$ in the low voltage range which is attributed to the electrochemical activity of \feS\ increases drastically while simultaneously a significant decrease in the antiperovskite-associated oxidation processes (O$_1$, O$_2$ and O$_*$) is observed. Both the appearance of an extra feature as well as observed irreversibility is similar to what is found in \LiFeSeO~\cite{singer2023elucidating}.

To determine the cut-off voltage at which the irreversible oxidation reaction O$_*$ set in, GCPL measurements using a constant current constant voltage protocol and a step-like increase of the maximum voltage in 0.1~V increments was performed (see Figs.~\ref{si-fig::stepgcpl}d,e,f). For all \LiFeSO\ materials under study, the threshold voltage beyond which the antiperovskite material deteriorates and \feS\ is produced amounts to 2.7~V, i.e., it is not affected by the above-discussed material changes due to heat treatment. As seen in the data, for all three materials we find an increase in the low voltage reactions as soon as the maximum voltage exceeds 2.6~V. The stable regime for electrochemical cycling is therefore restricted to a maximum voltage of 2.6~V.

\begin{figure}[h!]
\centering
\includegraphics[width=0.5\columnwidth]{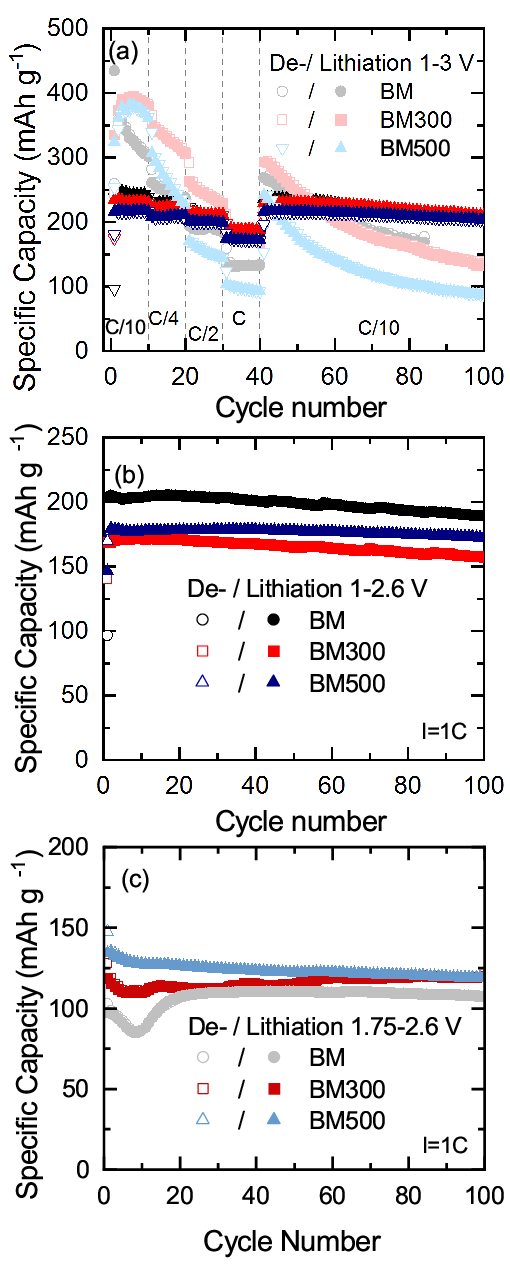}
    \caption{(a) Rate performance  of \lis-BM, -BM300 and -BM600 in the potential window 1-3~V and 1-2.6~V. Specific dis-/charge capacities of BM, -BM300 and -BM500 in the potential window (b) 1 to 2.6~V and (c)  between 1.75~ and 2.6~V at 1C. Note that the legend in (b) also describes data with the matching colour in (a).}
    \label{fig:EC}
\end{figure}

Performance characteristics of BM, BM300, and BM500 are presented in Fig.~\ref{fig:EC}a. To evaluate the impact of the high voltage anionic process on high-current capability and stability, two sets of measurements were performed: one extending up to 3~V and another one within the limited potential range of up to 2.6~V to avoid the anionic process O$_*$. Within the first cycles, two significant trends emerge. Firstly, when cycling up to 3~V, all studied materials exhibit an impressive initial capacity of approximately 400~\mahg\ at a C/10 current rate. This marks a notable achievement and unequivocally underscores the substantial advantages and potential offered by both the material and the chosen synthesis approach. Compared to the initial capacity achieved by previous measurements on lithium-rich antiperovskites cathodes of 285~\mahg\ for \LiFeSO\ by Mikhailova~\etal \cite{Mikhailova.2018}, 220~\mahg\ for (Li$_2$Fe$_{0.8}$Mn$_{0.2}$)SO by Deng~\etal \cite{Deng.2023-Li2FeMnSO},  245~\mahg\ for (Li$_2$Fe)S$_{0.7}$Se$_{0.3}$O by Mohamed~\etal \cite{Mohamed.2021},  164 \mahg\ / 300~\mahg\ for \LiFeSeO\ by Mohamed~\etal \cite{Li2FeSeO-2022} / Singer~\etal \cite{singer2023elucidating}, this is by far the highest capacity every achieved.
In contrast, in the constrained potential range, BM, BM300, and BM500 all show capacities around 220~\mahg. Secondly, the cycling dependence of all materials is rather similar in the restricted voltage regime but strongly differs in the extended regime. To be specific, when cycling up to 3~V, in the first nine cycles BM shows a significant loss of capacity, while BM300 and BM500 show an initial increase followed by a significant decrease (see Fig.~\ref{fig:EC}a). At higher currents, the differences between the different voltage ranges become even more pronounced: In the full range, we observe average capacities of 261, 188, and 132~\mahg\ at C/4, C/2, and 1~C and 230, 210, 180~\mahg\ in the limited range for BM. For B300, the same measurements yield 325, 240, and 175~\mahg\ (full range) and 220, 210, 187~\mahg\ (limited range), and for BM500 250, 150, and 95~\mahg\ (full range) and  205, 200, 172~\mahg\ (limited range). Overall, the limited potential range is clearly beneficial for all materials: First and foremost, it results in drastically improved cycling stability as well as in higher converted capacity at high currents.  We also particularly emphasize the very good high-current capability of the \LiFeSO\ samples in the limited range, which only show low reductions of around 20\% in capacity with a tenfold increase in current.

The outstanding electrochemical capabilities and stability achieved for \LiFeSeO\ when avoiding the high voltage anionic reaction are validated by galvanostatic cycling experiments conducted at a rate of 1C presented in Fig.\ref{fig:EC}b. Specifically, the initial capacity reaches approximately 200~\mahg\ in the case of BM, 175~\mahg\ for BM500, and 170~\mahg for BM300. The best cycling stability of the materials under study is found for BM500 which shows 96~\% capacity retention after 100 cycles. Note, that this is indeed the highest capacity retention ever reported for a lithium-rich antiperovskite material.

In order to quantify and separate the contributions of impurity phases to the observed reversible capacity, further studies with the minimum voltage limited to 1.75~V are shown in Fig.~\ref{fig:EC}c. In this potential range, we expect no electrochemical activity from impurity phases (active only below 1.75~V (see Fig.\ref{si-fig::stepgcpl})) so that the resulting capacity can be completely attributed to the \lis\ component.

Fig.~\ref{fig:EC}c shows that the capacity of the BM material is reduced by 100~\mahg\ when cycling in the regime 1.75 -- 2.6~V. This difference corresponds very well to the value obtained by direct discharge (see Fig.~\ref{si-fig::stepgcpl}a) which hence further evidences that impurity phases are electrochemically not active above 1.75~V. Note, that the actual capacity arising from the antiperovskite phase is higher than given in Fig.~\ref{fig:EC}c as the specific capacity is normalised to the total mass which, depending on the post-synthesis treatment, includes amounts of in this voltage window inactive impurity phases. The fact that the BM sample contains the highest proportion of now inactive foreign phase among the studied materials hence straightforwardly explains that its capacity in Fig.~\ref{fig:EC}c is lower than that of BM300 and BM500. In addition to this mass effect, our GCPL and CV studies in Figs.~\ref{si-fig::stepgcpl} and \ref{si-fig::cv-all} indicate slight but visible differences in kinetics between the samples so that the exact voltages of the relevant processes slightly differ, too. This also leads to small differences between the measured capacities of the samples.

The obtained capacities associated with the bare title compounds presented in Fig.~\ref{fig:EC}c show substantial changes to the overall capacity data in Fig.~\ref{fig:EC}b. Firstly, BM does not show the best performance anymore. After 100 cycles, BM shows a capacity of 110~\mahg, while BM300 and BM500 present more than 120~\mahg. In addition, there is a minimum in reversible capacity after a few ($\sim 10$) cycles. We attribute~\footnote{We however note that, for the BM sample, a clear separation of the electrochemical activity associated with impurities is more difficult than for the other materials due to the close proximity to the antiperovskite processes.} this behaviour to changing kinetics leading to a small contribution of the impurities (see Fig.~\ref{si-fig::Potential-profiles-1.6-2.7} in the Supplementary Information, where an initial increase of the voltage plateaus is visible).

In general, Fig.~\ref{fig:EC}c clearly shows that, in the first 100 cycles, the electrochemical performance of the bare antiperovskite phase is almost independent on the post-treatment method. The fact that the particle sizes of the materials strongly differ (sub to few micrometer-sized particles (BM) compared to particles above the ten-micrometre range (BM500) \cite{Mohamed.2023-synthese}) indicates high ionic conductivity of the \lis-phase and highlights its potential for lithium-ion battery application.

\begin{figure*}
\begin{center}
 \includegraphics[width=\textwidth]{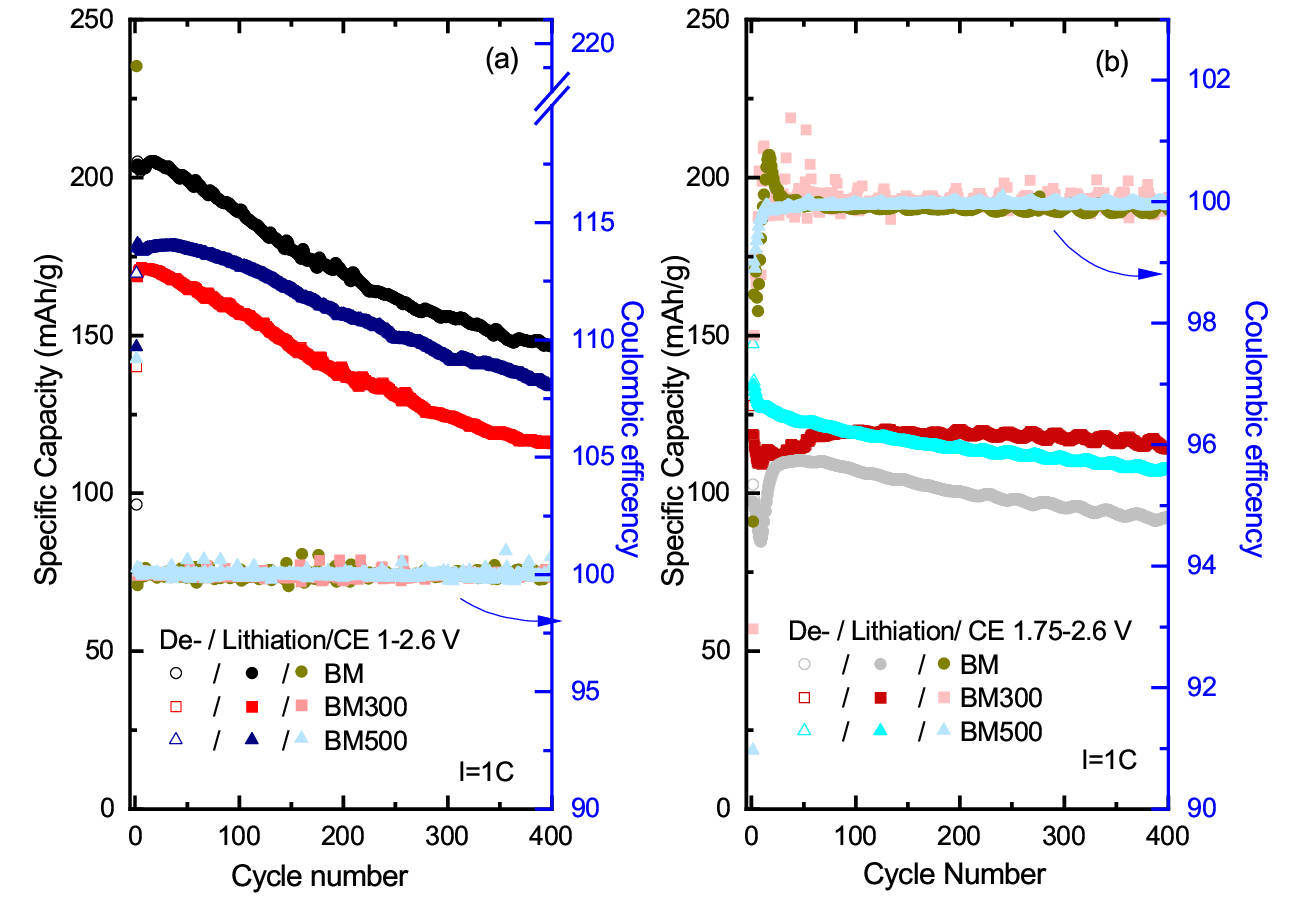}
    \caption{Specific charge/discharge capacities as well as Coulombic efficiency of BM, BM300 and BM500 electrodes at 1~C (a) in the voltage range 1 to 2.6~V and (b) 1.75 to 2.6~V.}
   \label{si-fig::GCPL+CE}
   \end{center}
\end{figure*}

Long-term galvanostatic cycling studies of BM, BM300 and BM500 in the restricted potential range from 1 to 2.6~V and from 1.75 to 2.6 V together with the Coulombic efficiency shown in Fig.~\ref{si-fig::GCPL+CE} in the SI give further insights into the practically accessible capacity. In the range from 1 to 2.6~V BM, BM300, and BM500 show a drop of 32\%, 38\%, and 29\% over 400 cycles. Here, we note the artificially high Coulombic efficiency (CE) of the BM sample of around 200~\% in the first cycle. This high CE is only possible due to the lithium metal counter electrode while in a full cell experiment the maximum achievable value in the following cycles is restricted to the discharge capacity of approximately 100~\mahg. Further restriction to only the antiperovskite processes (voltage range 1.75 to 2.6~V, Fig.~\ref{si-fig::Potential-profiles-1.6-2.7}b) leads to a discernible reduction in Coulombic efficiency (CE) to approximately 95~\%. Restricting the voltage window to 1.75 - 2.6~V therefore effectively minimises the loss of practically achievable capacity to around 5\% after the first cycle, compared to a 50~\% loss with the low voltage limit of 1 V (see Fig.~\ref{si-fig::GCPL+CE}a). At the same time, an improvement in cycle stability can be observed, as shown by the BM300 example, which has a loss of only 5\% in over 400 cycles compared to 38~\% in the extended voltage range 1-2.6~V. These results imply two major findings: (1) The cationic electrochemical processes of \lis\ are overall very stable. (2) Further performance improvements can be obtained by reducing the amount of impurities and the prevention of the conversion reaction O$_*$.

\section{Conclusions}

Our detailed electrochemical study on lithium-rich antiperovskite cathode materials allows us to separate cationic and anionic redox processes. With the high voltage anionic process an astonishing electrochemical capacity  of around 400~\mahg\ is initially reachable. Our results however also demonstrate that the high voltage anionic process is the root cause of the poor cycling stability reported in all previous studies on lithium-rich antiperovskites. By avoiding the so-called anionic process (O$_*$), a highly improved cycling stability is achieved. The success of this approach is proven with the BM500 sample which exhibits at 1~C only 4~\% capacity fading over 100 cycles at the remarkable capacity of about 175~\mahg. The data hence prove drastic performance improvements of \LiFeSO. While strongly differing particles size does not visibly affect the electrochemical performance of the materials under study, impurities are a key factor. In addition, by appropriately restricting the voltage regime we have also investigated the electrochemical performance of the bare antiperovskite phase, which in the restricted voltage regime provides over 400 cycles the stable capacity of about 110~\mahg. In summary, by separating cationic and anionic redox activity we have shown the consequences of each process as well as the implications on the overall electrochemical performance. Our data highlight the potential of the pure cationic electrochemical reaction in lithium-rich antiperovskites, thereby showing the route to further improve their performance in electrochemical energy storage.

\section{Additional Material}

\begin{figure*}
\begin{center}
 \includegraphics[width=\textwidth]{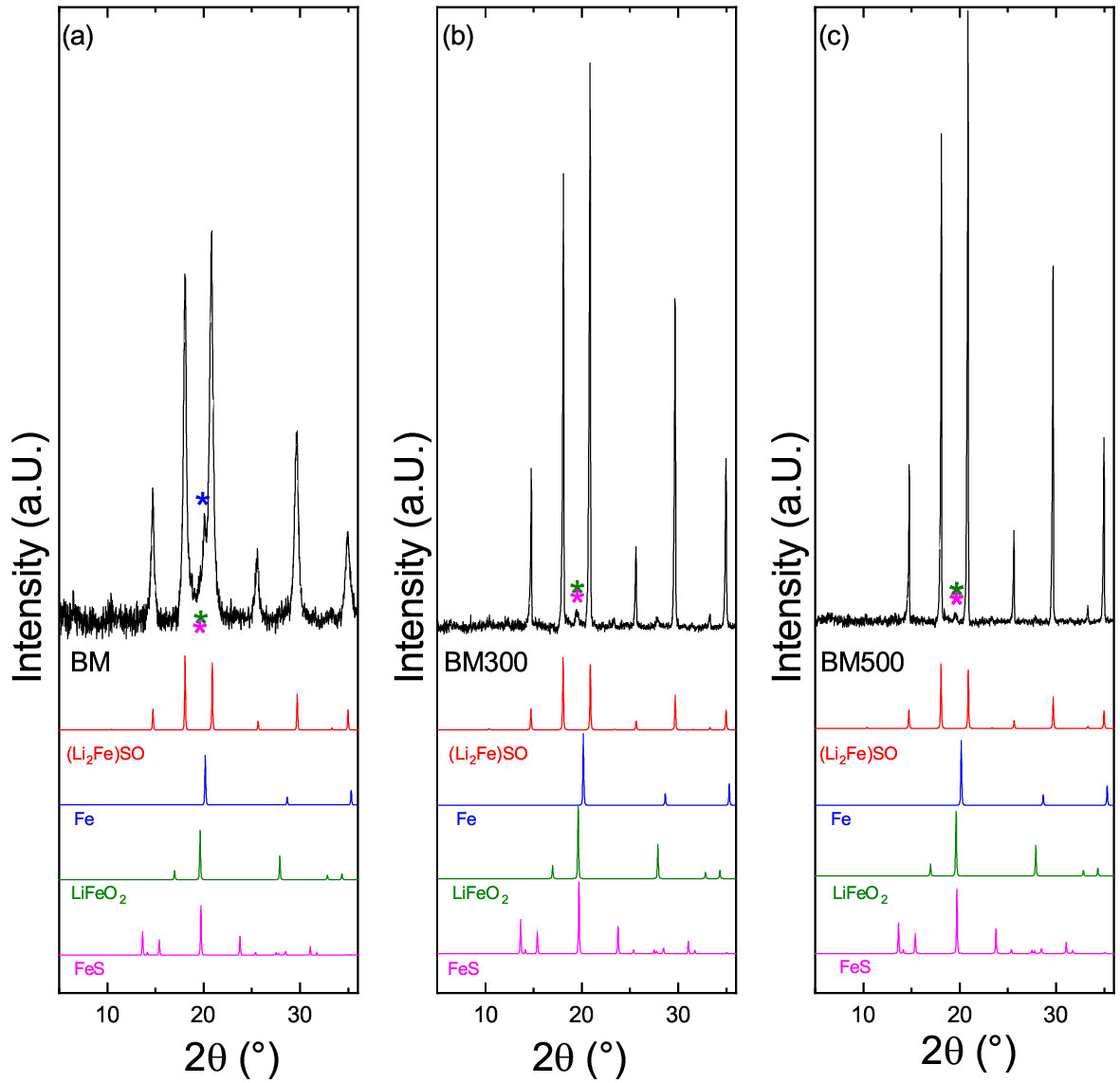}
    \caption{XRD patterns of BM (a), BM300 (b) and BM 500 (c) as well as reference patterns for \lis\ (ICSD No. 253936 \cite{Lai.2017}), Fe (ICSD No. 53451 \cite{Fe-Ref.1967}),  LiFeO$_2$ (ICSD No. 155031 \cite{VucinicVasic.2006-LiFeO2-Ref}) and FeS (ICSD No. 53526 \cite{Coey.1979-FeS-53526}). }
   \label{si-fig::XRD}
   \end{center}
\end{figure*}

\begin{figure*}
\begin{center}
 \includegraphics[width=\textwidth]{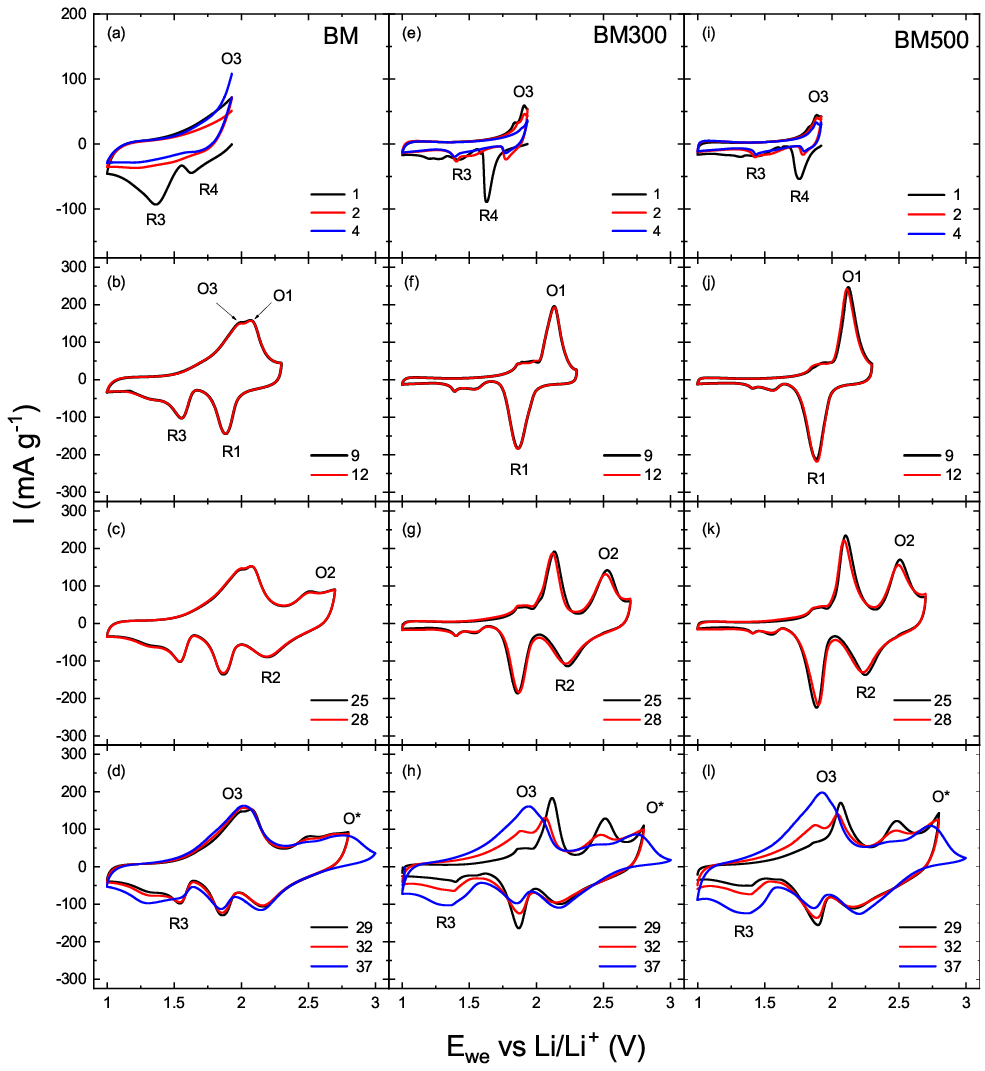}
    \caption{Cyclic voltammogramms of BM (left column), BM300 (middel column) and  BM500 (right column) in a potential window from 1~V  to OCV (a,e,i), 2.3~V (b,f,j), 2.7~V (c,g,k) and up to 3~V (d,h,l). The cyclic voltammograms for all post-heated materials (BM300 and BM500) show a clear increase in the low-voltage function O3/R3 and a decrease in O1,O2 only after the voltage range has been extended to the high-voltage process O*.}
   \label{si-fig::cv-all}
   \end{center}
\end{figure*}

\begin{figure*}
\begin{center}
 \includegraphics[width=\textwidth]{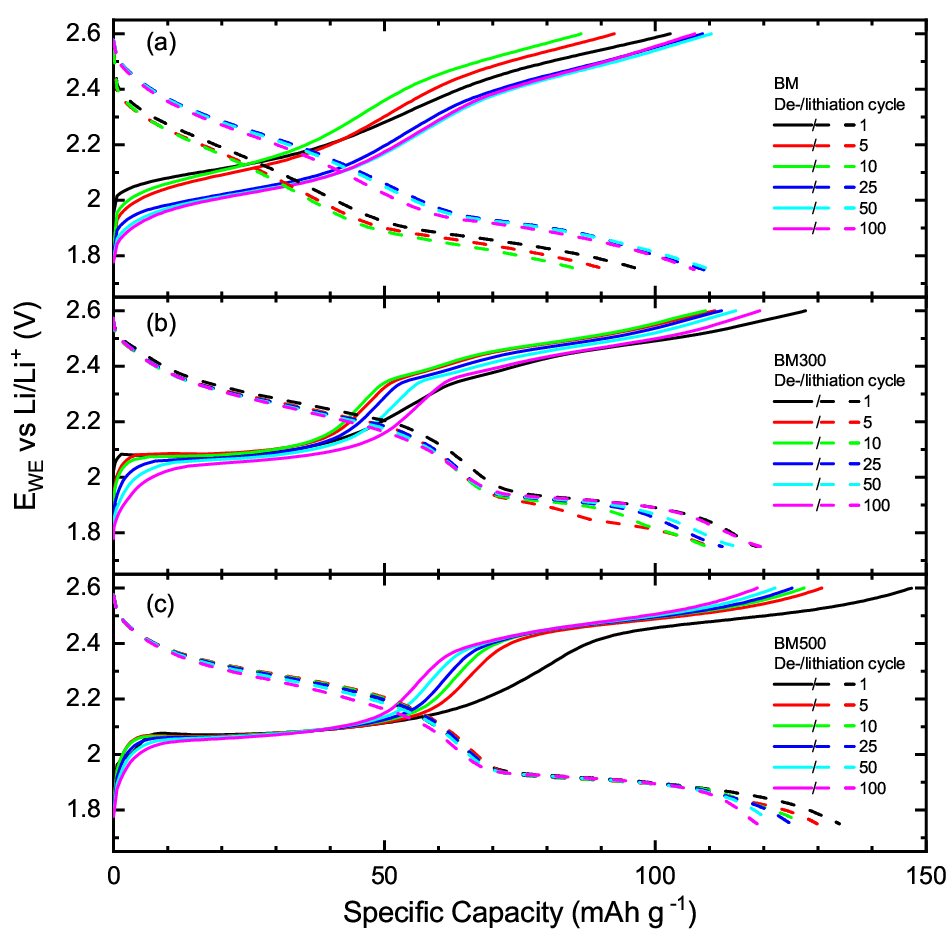}
    \caption{Potential profiles of BM (a), BM300 (b), BM500 (c) in the voltage range of 1.75 to 2.6~V.}
   \label{si-fig::Potential-profiles-1.6-2.7}
   \end{center}
\end{figure*}

\section*{Acknowledgements}
Financial support by Deutsche Forschungsgemeinschaft (DFG) through projects KL 1824/20-1 and GR 5987/2-1 is acknowledged. Work has also been supported within the framework of the Excellence Strategy of the Federal and State Governments of Germany via Heidelberg University's flagship EMS initiative. M.A.A. Mohamed thanks the IFW excellence program for financial support.

\balance

\bibliography{literatur.bib}

\end{document}